\font\bfs=cmb10 at 7pt

\font\bb=msbm10 at 10pt

\font\cal=cmsy10 at 9pt
\font\cals=cmsy10 at 8pt

\font\rms=cmr10 at 8pt

\font\sf=cmss10 at 10pt 

\def\0#1{\mbox{\rm#1}}
\def\1#1{\mbox{\bb#1}}
\def\2#1{\mbox{\bf#1}}
\def\3#1{{\cal #1}}
\def\4#1{\mbox{\cals#1}}
\def\5#1{\mbox{\sf#1}}
\def\6#1{\mbox{\rms #1}}
\def\7#1{\mbox{\bfs #1}}
\def\8#1{{\tilde #1}}
\def\9#1{{\breve #1}}

\def\BEq{\begin{equation}}
\def\EEq{\end{equation}}
\def\BEqA{\begin{eqnarray}}
\def\EEqA{\end{eqnarray}}
\def\BEn{\begin{enumerate}}
\def\EEn{\end{enumerate}}

\def\Cbb{\mbox{\bb C}}

\def\Rbb{\mbox{\bb R}}

\def\tav{\hbox{
\kern-1pt\rule[0pt]{1.5pt}{.8pt}{\kern-3.3pt}
\rule[0pt]{.4pt}{5pt}{\kern-4.4pt}
\rule[5pt]{4.5pt}{.8pt}{\kern-3.45pt}
\rule[0pt]{.4pt}{5.3pt}{\kern-1pt}
}}

\def\Cc{{\cal C}}

\def\dag{\dagger}

\def\adj{{^{\dag}}}

\def\from{\kern-2pt\leftarrow\kern-2pt}

\def\bra{\langle}

\def\ket{\rangle}

\def\II{|\kern-1pt |}

\def\Cliff{\mathop{\hbox{\rm Cliff}}\nolimits}

\def\tav{\hbox{
\kern-1.0pt
\rule[0pt]{1.3pt}{.8pt}{\kern-3.6pt}
\rule[0pt]{.4pt}{6pt}{\kern-3.0pt}
\rule[4.5pt]{3.0pt}{.8pt}{\kern-3.3pt}
\rule[0pt]{.4pt}{5pt}{\kern-1pt}
}}

\def\adj{{^{\dag}}}

\def\dag{\dagger}

\def\from{\kern-2pt\leftarrow\kern-2pt}

\def\lmult{{\lfloor\kern-5pt\lfloor}}
\def\rmult{{\rfloor\kern-5pt\rfloor}}

\def\adj{{^{\dag}}}
\def\dag{\dagger}

\def\bra{\langle}

\def\ket{\rangle}

\def\II{|\kern-1pt |}

\documentclass{article}

\usepackage{makeidx,graphics,latexsym}

\begin{document}

\title{{\bf Clifford statistics and the temperature limit in
the theory of fractional quantum Hall effect}}

\author{
Andrei A.
Galiautdinov \\
\normalsize {\it School of Physics, Georgia Institute of
Technology, Atlanta, Georgia 30332-0430 }}

\date{\today}
\maketitle

\abstract{Using the recently discovered Clifford statistics
we propose a simple model for the grand canonical ensemble
of the carriers in the theory of fractional quantum Hall
effect. The model leads to a
temperature limit associated with the
permutational degrees of freedom of such an ensemble. 
 We also relate Schur's theory of projective
representations of the permutation groups to physics, and
remark on possible extensions of the second
quantization procedure.}

\vskip20pt


\vskip20pt


In a series of papers, building on the work on
nonabelions of Read and Moore
\cite{moore90, read92}, Nayak and Wilczek
\cite{NW, WILCZEK, wilczek} (see also
\cite{Wilczek82} on how spinors can describe aggregates)
proposed a startling new spinorial statistics for the
fractional quantum Hall effect (FQHE) carriers. The
prototypical example is furnished by a so-called Pfaffian
mode (occuring at filling fraction $\nu = 1/2$), in which
$2n$ quasiholes form an
$2^{n-1}$-dimensional irreducible multiplet of the
corresponding braid group. The new
statistics is clearly non-abelian: it represents the
permutation group
$S_{N}$ on the $N$ individuals by a non-abelian group of
operators in the $N$-body Hilbert space, a projective
representation of $S_{N}$.

We have undertaken a systematic study of this
statistics elsewhere, aiming primarily
at a theory of elementary processes in quantum theory of
space-time. We have called the new statistics {\it Clifford},
to emphasize its intimate relation to Clifford algebras and
projective representations of the permutation
groups. The reader is referred to \cite{fs, FG, GF} for
details.

Since the subject is new, many unexpected effects in the
systems of particles obeying Clifford statistics may arise in
future experiments. One simple effect, which seems
especially relevant to the FQHE, might be observed in a grand
canonical ensemble of Clifford quasiparticles. In this
paper we give its direct derivation first.

Following Read and
Moore \cite{read92} we postulate that only two quasiparticles
at a time can be added to (or removed from) the FQHE
ensemble. Thus, we start with an $N=2n$-quasiparticle
effective Hamiltonian whose only relevant to our
problem energy level
$E_{2n}$ is
$2^{n-1}$-fold degenerate. 
The degeneracy of the ground mode with no quasiparticles
present is taken to be
$g(E_0)=1$.

Assuming that
adding a {\it pair} of quasiparticles to the composite
increases the total energy by $\varepsilon$, and ignoring
all the external degrees of freedom, we can tabulate the
resulting many-body energy spectrum as follows:
\BEq
\begin{array}{|l|c c c c c c c c|}\hline
{\rm Number \, of \, Quasiparticles}, \, N= 2n   &  0 & 2 & 4
& 6 & 8 & 10 & 12 & \cdots \\ \hline
{\rm Degeneracy}, \, g(E_{2n}) = 2^{n-1} & 0 & 1 & 2 & 4 & 8
& 16 & 32 &
\cdots \\ \hline
     {\rm Composite \, Energy}, \,E_{2n}   &  0\varepsilon &
1\varepsilon & 2\varepsilon & 3\varepsilon & 4\varepsilon &
5\varepsilon & 6\varepsilon &  \cdots \\ \hline
\end{array}
     \EEq

Notice that the energy levels so defined furnish irreducible
multiplets for projective representations of permutation
groups in Schur's theory \cite{schur}, as was first pointed
out by Wilczek \cite{WILCZEK, wilczek}.

We now consider a grand canonical ensemble of Clifford
quasiparticles.

     The probability that the composite
contains $n$ pairs of quasiparticles, is
     \begin{eqnarray}
     P(n,T)=
     \frac{ g(E_{2n})e^{(n\mu-E_{2n})/k_BT} }{ 1 +
\sum_{n=1}^
     {\infty} g(E_{2n})e^{(n\mu-E_{2n})/k_BT} } \nonumber  \\
     \equiv \frac{ 2^{n-1} e^{-(n\mu-E_{2n}) /k_BT} }
     { 1 + \sum_{n=1}^{\infty}2^{n-1} e^{-n(\varepsilon -\mu)
/k_BT} },
\end{eqnarray}
where $\mu$ is the quasiparticle chemical potential.
     The denominator of this expression is the grand
partition function of the composite,
     \begin{eqnarray}
     Z(T)= 1 + \sum_{n=1}^{\infty}
g(E_{2n})e^{(n\mu -E_{2n})/k_BT} \nonumber \\
     \equiv 1 + \sum_{n=1}^{\infty}2^{n-1}
	 e^{-n(\varepsilon -\mu) /k_BT}
     \end{eqnarray}
     at temperature $T$.

     Now,
     \begin{eqnarray}
\sum_{n=1}^{\infty}2^{n-1}e^{-nx}=&e^{-x} \;[2^0 e^{- 0x}
     +2^1 e^{-1x} +2^2e^{-2x}+\cdots] \cr
     =& e^{-x} \; \sum_{n=0}^{\infty}e^{n(\ln2-x)}.
     \end{eqnarray}

     The partition function can therefore be written as
     \begin{eqnarray}
      Z(T) =1 + e^{- (\varepsilon -\mu) /k_BT}
     \sum_{n=0}^{\infty}e^{n(\ln2-(\varepsilon -\mu) /k_BT)}.
     \end{eqnarray}

  This leads to two interesting possibilities (assuming
$\varepsilon > \mu$):

     1) Regime $0<T<T_c$, where
     \begin{equation}\label{eq:TC}
     T_c= \frac{\varepsilon -\mu}{k_B\ln{2}}\/.
     \end{equation}

     Here the geometric series converges and
     \begin{eqnarray}
     Z(T) = \frac{ 1 - e^{- (\varepsilon -\mu) /k_BT} }
     { 1 - 2e^{- (\varepsilon -\mu)/k_BT} } =
\frac{ 2e^{-(\varepsilon -\mu)/2k_BT} }
{1-2e^{-(\varepsilon -\mu)/k_BT}} \; {\rm
sinh}\left( \frac{ \varepsilon -\mu}{ 2k_BT }\right).
     \end{eqnarray}

     The probability distribution is given by
     \begin{eqnarray}
     P(n, T) = \frac{  2^{n-1}e^{-n (\varepsilon -\mu) /k_BT}
     (1 - 2e^{- (\varepsilon -\mu)/k_BT})} {1 - e^{-
(\varepsilon -\mu) /k_BT} }.
     \end{eqnarray}

     2) Regime $T\geq T_c$.

     Under this condition the partition function diverges:
     \begin{equation}
     Z(T) = +\infty,
     \end{equation}
     and the probability distribution vanishes:
     \begin{equation}
     P(n,T) = 0 .
     \end{equation}

     This result indicates that the temperature $T_c$ of
(\ref{eq:TC}) is the upper bound of the intrinsic
temperatures that the
    quasiparticle ensemble can have. Raising the
temperature brings the system to higher energy levels which
are more and more degenerate, resulting in a heat capacity
that diverges at the temperature $T_c$\/.

To experimentally observe this
effect, a FQHE system should be subjected to a condition
where quasiparticles move freely between the specimen
and a reservoir, without exciting other degrees of freedom
of the system.

     A similar limiting temperature phenomenon seems to
occur in nature as
     the Hagedorn limit in particle physics \cite{hagedorn}.

Knowing the partition function allows us to find various
thermodynamic quantities of the quasiparticle system
for sub-critical temperatures $0<T<T_c$. We are particularly
interested in the average number of {\it pairs}
in the grand ensemble:
     \begin{equation}
     \label{eq:av.number}
     \bra n \ket_{\rm{Cliff}}= \lambda \frac{\partial \ln
Z}{\partial
\lambda},
     \end{equation}
     where $\lambda = e^{\mu /k_BT}$, or after
some algebra,
     \begin{eqnarray}
     \langle n(T) \rangle_{\rm{Cliff}} = \frac{e^{-
(\varepsilon -\mu) /k_BT}}
     {(1 - e^{-(\varepsilon -\mu)/k_BT})(1-2e^{-
(\varepsilon -\mu) /k_BT})} .
     \end{eqnarray}

     We can compare this with the familiar
Bose-Einstein,
     \begin{equation}
     \langle n(T) \rangle_{\rm{BE}} =
     \frac{1}{e^{(\varepsilon -\mu) /k_BT}-1} \equiv
     \frac{e^{- (\varepsilon -\mu) /k_BT}}{1 - e^{-
(\varepsilon -\mu) /k_BT}},
     \end{equation}
and Fermi-Dirac,
     \begin{equation}
     \langle n(T) \rangle_{\rm{FD}} =
     \frac{1}{e^{(\varepsilon -\mu) /k_BT}+1} \equiv
    \frac{e^{- (\varepsilon -\mu) /k_BT}}{1 + e^{-
(\varepsilon -\mu) /k_BT}},
     \end{equation}
distributions.
For the Clifford oscillator,
$\langle n(T) \rangle_{\rm{Cliff}} \rightarrow +\infty$ as
$T \rightarrow T_c -$, as had to be expected.

Let us now turn to projective representations of the
symmetric (permutation) groups that have long been known to
mathematicians, but received little attention from
physicists. Such representations were overlooked in physics
much like projective representations of the rotation groups
were overlooked in the early days of quantum mechanics.

For convenience, following \cite{schur,
KARPILOVSKY, HOFFMAN} ({\it cf.} also \cite{WILCZEK,
wilczek}), we briefly
recapitulate the main results of Schur's theory.

One especially useful presentation of the
symmetric group $S_N$ on $N$ elements is given by
\BEqA
S_N &=& \bra \; t_1, \dots  , t_{N-1} \; : \; t_i^2 = 1,\;
(t_jt_{j+1})^3=1, \; t_kt_l=t_l t_k \; \ket ,\nonumber \\
&& 1\leq i
\leq N-1,\; 1\leq j \leq N-2, \; k\leq l-2 .
\EEqA
Here $t_i$ are transpositions,
\BEq
t_1=(12), t_2=(23), \dots , t_{N-1}=(N-1 N).
\EEq
Closely related to $S_N$ is the group $\tilde S_N$,
\BEqA
\label{eq:REPRESENTATIONGROUP}
\tilde S_N &=& \bra \; z, {t'}_1, \dots  , {t'}_{N-1}\; :\;
z^2=1,\; z{t'}_i = {t'}_i z,\;{t'_i}^2 = z,\;
({t'}_j{t'}_{j+1})^3=z, \; {t'}_k{t'}_l= z \, {t'}_l {t'}_k
\;
\ket ,\nonumber
\\ && 1\leq i
\leq N-1,\; 1\leq j \leq N-2, \; k\leq l-2 .
\EEqA

A celebrated theorem of Schur (Schur, 1911 \cite{schur})
states the following:

\noindent (i) The group $\tilde S_N$ has order $2(n!)$.

\noindent (ii) The subgroup $\{1,z\}$ is central, and is
contained in the commutator subgroup of $\tilde S_N$,
provided
$n\geq 4$.

\noindent (iii) $\tilde S_N / \{1,z\} \simeq S_N$.

\noindent (iv) If $N < 4$, then every projective
representation of
$S_N$ is projectively equivalent to a linear representation.

\noindent (v) If $N\geq 4$, then every projective
representation of
$S_N$ is projectively equivalent to a representation
$\rho$,
\BEqA
\rho(S_N) &=& \bra \; \rho(t_1), \dots ,
\rho(t_{N-1}): \rho(t_i)^2 = z, (\rho(t_j)\rho(t_{j+1}))^3=z,
\nonumber \\
&& \rho(t_k)\rho(t_l)= z \, \rho(t_l) \rho(t_k)
\; \ket ,
\nonumber \\ && 1\leq i
\leq N-1,\; 1\leq j \leq N-2, \; k\leq l-2 ,
\EEqA
where $z = \pm 1$. In the case $z = +1$, $\rho$
is a linear representation of $S_N$.

The group $\tilde S_N$ (\ref{eq:REPRESENTATIONGROUP}) is
called the {\it representation group} for $S_N$.

The most elegant way to construct a {\it projective}
representation
$\rho(S_N)$ of $S_N$ is by using the complex Clifford
algebra
${\Cliff}_{\Cbb}(V, g)\equiv \Cc_N$ associated with the 
real vector space
$V=N\Rbb $,
\BEqA
\label{eq:CLIFFALGEBRA}
\{\gamma_i, \gamma_j\}=-2g(\gamma_i, \gamma_j).
\EEqA
Here $\{\gamma_i\}_{i=1}^N$ is an orthonormal basis
of $V$ with respect to the symmetric bilinear form
\BEq
g(\gamma_i, \gamma_j)=+\delta_{ij}.
\EEq
Clearly, any subspace $\bar V$ of $V=N\Rbb $ generates a
subalgebra
${\Cliff}_{\Cbb}(\bar V,\bar g)$, where $\bar g$ is the
restriction of
$g$ to $\bar V \times \bar V$. A particularly
interesting case is realized when $\bar V$ is
\BEq
\bar V:=\left\{ \sum_{k=1}^N \alpha^k \gamma_k \; : \;
\sum_{k=1}^N
\alpha_k=0\right\}
\EEq
of codimension one, with the
corresponding subalgebra denoted by
$\bar \Cc_{N-1}$ \cite{HOFFMAN}.

If we consider a special basis $\{t'_k\}_{k=1}^{N-1}\subset
\bar V$ (which is {\it not} orthonormal) defined by
\BEq
\label{eq:SWAPS}
t'_k:=\frac{1}{\sqrt{2}}(\gamma_k - \gamma_{k+1}), \quad
k=1, \dots, N-1,
\EEq
then the group generated by this basis is isomorphic to
$\tilde S_N$. This can be seen by mapping
$t_i$ to
${t'}_i$ and $z$ to -1, and by noticing that

\noindent 1) For $k=1, \dots, N-1$:
\BEq
{t'_k}^2 
= - 1;
\EEq

\noindent 2) For $N-2 \geq j$:
\BEq
(t'_jt'_{j+1})^3
= -1 ;
\EEq

\noindent 3) For $N-1\geq m>k+1$:
\BEq
t'_kt'_m = - t'_mt'_k ,
\EEq
as can be checked by direct calculation.

One choice for the matrices is provided by the following
construction (Brauer and Weyl, 1935 \cite{brauer}):
\begin{eqnarray}
\label{eq:BRAUERWEYL}
\gamma_{2k-1} = \sigma_3 \otimes \cdots \otimes \sigma_3
\otimes
(i\sigma_1) \otimes {\bf 1} \otimes \cdots \otimes {\bf 1},
\cr \gamma_{2k} =
\sigma_3 \otimes \cdots \otimes \sigma_3 \otimes
(i\sigma_2) \otimes {\bf 1} \otimes \cdots \otimes {\bf 1}, \cr
i = 1, \, 2, \, 3,..., \, M,
\end{eqnarray}
for $N=2M$. Here $ \sigma_1$, $ \sigma_2$ occur in the
$k$-th position, the product involves $M$ factors, and
$\sigma_1$,
$\sigma_2$, $\sigma_3$ are the Pauli matrices.

If $N=2M+1$,
we first add one more matrix,
\BEq
\gamma_{2M+1} = i\,\sigma_3 \otimes \cdots \otimes \sigma_3
\quad (M \; {\rm factors}),
\EEq
and then define:
\begin{eqnarray}
\Gamma_{2k-1} &:=& \gamma_{2k-1}\oplus \gamma_{2k-1},
\nonumber \\
\Gamma_{2k} &:=&
\gamma_{2k}\oplus \gamma_{2k},
\nonumber \\
\Gamma_{2M+1} &:=&
\gamma_{2M+1}\oplus (-\gamma_{2M+1}).
\end{eqnarray}

The representation $\rho(S_N)$ so constructed is
reducible. An irreducible module of $\bar\Cc_{N-1}$
restricts that representation to the irreducible
representation of
$\tilde S_N$, since
$\{t'_k\}_{k=1}^{N-1}$ generates $\bar \Cc_{N-1}$ as an
algebra
\cite{HOFFMAN}.

To relate Schur's theory to physics we may try to define a
new, purely permutational variable of the Clifford composite,
whose spectrum would reproduce the degeneracy of Read and
Moore's theory.

A convenient way to define such a variable
is by the process of {\it quantification} (often called
second quantization), which is used in all the usual
quantum statistics --- by mapping the one-body Hilbert space
into a many-body operator algebra. This procedure was
described in detail in
\cite{FG}.

Let us thus assume that if there is just one quasiparticle in
the system, then there is a limit on its localization, so
that the quasiparticle can occupy only a finite number of
sites in the medium, say $N=2n$. We further assume that the
Hilbert space of the quasiparticle is {\it real} and
$N=2n$-dimensional, and that a one-body
variable (which upon quantification corresponds to the
permutational variable of the ensemble) is an antisymmetric
generator of an orthogonal transformation of the form
\begin{eqnarray}
\label{eq:G}
G:= A
\left[\matrix{{\bf 0}_n & {\bf 1}_n \cr
-{\bf 1}_n& \hphantom{-}{\bf 0}_n}\right] ,
\end{eqnarray}
where $A$ is a constant coefficient. Note that in the complex
case this operator would be proportional to the imaginary unit
$i$, and the corresponding unitary transformation would be a
simple multiplication by a phase factor with no
observable effect. Since the quantified operator algebra for
$N>1$ quasiparticles will be complex, the effect of just one
such ``real'' quasiparticle should be regarded as
negligible in the grand canonical ensemble.

In the non-interacting case the process of quantification
converts
$G$ into a many-body operator $\hat G$ by the rule
\begin{eqnarray}
\label{eq:hat G}
\hat{G}&:= &\sum_{l, j}^N
\hat e_l G^l{}_j \hat e^j,
\end{eqnarray}
where usually $\hat e_i$ and $\hat e^{j}$ are
creators and annihilators, but in more general situations
are the generators (that appear in the commutation or
anticommutation relations) of the many-body operator
algebra. If $M_{ij}$ is the metric (not necessarily
positive-definite) on the one-body Hilbert space then
\BEq
\label{eq:conditionGENERATORS}
\hat{e}_i^{\dag} = M_{ij} \hat e^{j}.
\EEq
In the
positive-definite case, $M_{ij} =\delta_{ij}$ with
$\hat{e}_i^{\dag} = \hat e^{i}$, as usual.

In Clifford statistics \cite{FG} the generators of the
algebra are Clifford units $\gamma_i=2\,\hat e_i
=-\gamma_i\adj$, so it is natural to assume that
quantification of $G$ proceeds as follows:
\begin{eqnarray}
\label{eq:HATG}
\hat{G}
&= & -A \sum_{k=1}^{n}
(\hat e_{k+n}\hat e^k -
\hat e_k\hat e^{k+n}) \cr
&= & +A \sum_{k=1}^{n}
(\hat e_{k+n}\hat e_k -
\hat e_k\hat e_{k+n}) \cr
&= & 2 A \sum_{k=1}^{n}\hat e_{k+n}\hat e_k \cr
&\equiv &
\frac{1}{2} A \sum_{k=1}^{n} \gamma_{k+n}
\gamma_k.
\end{eqnarray}

By Stone's theorem, the generator $\hat G$ acting on the
spinor space of the complex Clifford composite of $N=2n$
individuals can be factored into a Hermitian operator
$\tilde O$ and an imaginary unit
$i$ that commutes strongly with $\tilde O$:
\BEq
\hat G = i \tilde O.
\EEq
We suppose that $\tilde O$ corresponds to the permutational
many-body variable mentioned above, and seek its spectrum.

We note that $\hat G$ is a sum of $n$
commuting anti-Hermitian algebraically independent operators
$\gamma_{k+n} \gamma_k, \; k=1,2,...,n$,
$(\gamma_{k+n} \gamma_k)^{ \dagger }=-\gamma_{k+n}
\gamma_k$,
$(\gamma_{k+n} \gamma_k)^2=-1$.
If we now use $2^n \times 2^n$
complex matrix representation of Brauer and Weyl
(\ref{eq:BRAUERWEYL}) for the
$\gamma$-matrices, we can simultaneously
diagonalize the $2^n\times2^n$ matrices representing the
commuting operators $\gamma_{k+n} \gamma_k$, and use their
eigenvalues, $\pm i$, to find the spectrum of
$\hat G$, and consequently of $\tilde O$. The final result
is obvious: there are $2^{2n}$ eigenkets of $\tilde
O$, corresponding to the dimensionality of the spinor
space of 
${\Cliff}_{\Cbb}(2n)$. In the irreducible
representation of
$S_N$ this number reduces to $2^{n-1}$, as
required by Read and Moore's theory.

Note that in this approach the possible number of the
quasiholes in the ensemble is fixed by the number of
the available sites, $N=2n$. A change in that
number must be accompanied by a change in the
dimensionality of the one-quasiparticle Hilbert space.
It is natural to assume that variations in the physical
volume of the entire system would privide such a mechanism. 

With regard to the quantification procedure
(\ref{eq:conditionGENERATORS}, \ref{eq:HATG}) mentioned
above, we point out that {\it a priori} there is no
compelling reason for using only the formalism of creation
and annihilation operators in setting up a many-body theory.

For example, if we choose to work exclusively with an
$N$-body system, then all the initial and final
selective actions (projections, or yes-no experiments) on
that system can be taken as {\it simultaneous} sharp
production and registration of {\it all} the
$N$ particles in the composite with no need
for one-body creators and annihilators. The theory would
resemble that of just one particle.  The elementary
non-relativistic quantum theory of atom provides such an
example.

Of course in real experiments much more
complicated processes occur. The number of particles
in the composite may vary, and if a special {\it vacuum}
mode is introduced, then those processes can conveniently be
described by postulating elementary operations of
one-body creation and annihilation. Using just the notion of
the vacuum mode and a simple rule by which the creation
operators act on the many-body modes, it is possible to show
(Weinberg \cite{WEINBERG1}) that {\it any} operator of
{\it such} a many-body theory may be expressed as a sum of
products of creation and annihilation operators.

In physics shifts in description are very frequent,
especially in the theory of solids. The standard example
is the phonon description of collective excitations in
crystal lattice. There the fundamental system is an ensemble
of a fixed number of ions without any special vacuum mode.
An equivalent description is in
terms of a variable number of phonons,
their creation and annihilation operators, and the
vacuum.

It is thus possible that a deeper theory underlying the
usual physics might be based on a completely new kind of
description. Finkelstein some time ago
\cite{DRFPRIVATE} suggested that the role of atomic
processes in such a theory might be played by swaps (or
permutations) of quantum space-time events. Elementary
particles then would be the excitations of a more
fundamental system. The most natural choice for the swaps is
provided by the differences of Clifford units
(\ref{eq:SWAPS}) defined above.

All that prompted us to generalize from the common statistics
to more general statistics, as was done in \cite{FG}. There,
the quantification rule (\ref{eq:conditionGENERATORS}) is a
consequence of the so-called representation principle.

\section*{Acknowledgements}

This work was aided by discussions with James Baugh and
David Finkelstein. It was partially supported by the M. and
H. Ferst Foundation.

\end{document}